\newif\ifws
\ifws

\documentclass{article}

\usepackage{graphicx}        
\usepackage[dvipsnames]{xcolor}

\usepackage{hyperref}
\hypersetup{
    colorlinks,
    linkcolor={blue!80!black},
    citecolor={red!75!black},
    urlcolor={blue!80!black}
}

\title{From Unitarity to Irreversibility: The Role of Infinite Tensor Products and Nested Wigner's Friends}

\author{Karl Svozil \\
        Institute for Theoretical Physics,
TU Wien,  \\
Wiedner Hauptstrasse 8-10/136,
1040 Vienna,  Austria
        }

\date{\today}
\begin{document}

\maketitle

\begin{abstract}
\end{abstract}

\else
\PassOptionsToPackage{dvipsnames}{xcolor}
\documentclass[
reprint,
 showpacs,
 showkeys,
 preprintnumbers,
 amsmath,amssymb,
 aps,
  pra,
  longbibliography,
 floatfix,
 ]{revtex4-2}

\usepackage{mathptmx}

\usepackage{amssymb,amsthm,amsmath}

\usepackage{mathbbol}

\usepackage{tikz}
\usetikzlibrary{calc,math}
\usepackage {pgfplots}
\pgfplotsset {compat=1.8}
\usepackage{graphicx}

\usepackage{xcolor}

\usepackage{hyperref}
\hypersetup{
    colorlinks,
    linkcolor={blue},
    citecolor={red!75!black},
    urlcolor={blue}
}

\usepackage{soul}
\usepackage{tcolorbox}
\tcbuselibrary{breakable} 

\begin{document}

\title{From Unitarity to Irreversibility: The Role of Infinite Tensor Products and Nested Wigner's Friends}

\author{Karl Svozil}
\email{svozil@tuwien.ac.at}
\homepage{http://tph.tuwien.ac.at/~svozil}

\affiliation{Institute for Theoretical Physics,
TU Wien,
Wiedner Hauptstrasse 8-10/136,
1040 Vienna,  Austria}

\date{\today}

\begin{abstract}
The transition from  unitary, reversible von Neumann-Everett quantum processes to non-unitary, irreversible processes and measurements is explored through infinite tensor products interpreted as nested, chained, or iterated Wigner's friend scenarios. Infinite tensor products can disrupt unitary equivalence through sectorization and factorization, drawing parallels to concepts from real analysis, recursive mathematics, and statistical physics.
\end{abstract}


\maketitle

\newpage
\fi

\section{Introduction}

Unitary equivalence can be surpassed through infinite means.
This observation is consistent with findings in number theory and analysis, where finite operations on rational numbers cannot yield results beyond the rational domain.
However, when infinite methods and techniques are employed---such as Cantor's diagonalization~\cite{cantor-set-engl,Yanofsky-BSL:9051621} or the construction of Specker sequences~\cite{specker57,specker-ges,kreisel}, exemplified by Chaitin's halting probability~\cite{chaitin3,calude-dinneen06}---it
becomes possible to conceptualize irrational, incomputable, and random (algorithmically incompressible) numbers.

The central problem this paper addresses is the quantum measurement problem: how can the non-unitary, irreversible measurement process (von Neumann's ``process 1'') emerge from the purely unitary and reversible evolution described by the Schr\"odinger equation (``process 2'')? Mainstream approaches to this problem include decoherence-based accounts, which explain the apparent collapse as a result of entanglement with the environment; objective collapse theories, which modify quantum dynamics to include a physical collapse mechanism; and Everettian or Many-Worlds interpretations, which posit that all outcomes occur in different branches of a universal wavefunction. This paper explores an alternative perspective, investigating the hypothesis that the transition to irreversibility is an emergent phenomenon that arises in the mathematical limit of infinitely complex systems. This limit is modeled using infinite tensor products, with the nested Wigner's Friend scenario serving as a conceptual framework for such an infinite regression. The advantage of this formalism is its ability to mathematically break unitary equivalence without altering the fundamental quantum postulates for finite systems.

The unitary group, which formalizes quantum state evolution (excluding irreversible measurements and processes such as tracings), is, like all groups, inherently `hermetic' by definition. In particular, its closedness under unitary transformations reflects a fundamental property of group theory, connected to mere permutations or one-to-one transformations of the identity element. Consequently, it seems that irreversibility cannot emerge from purely unitary evolution.
To explore this further, let us revisit the historical arguments that gave rise to the conundrum posed by von Neumann and others.


In the von Neumann scheme for ideal quantum measurement~\cite{landau-1927,wigner-1963},
the `object' is prepared in an initial state $\vert \psi \rangle$.
With respect to a `mismatching' context (relative to that preparation)---or,
 equivalently, orthonormal basis or maximal operator~\cite[Satz~8]{v-neumann-31} (\cite[Theorem~1, \S~84]{halmos-vs})---$\vert \psi \rangle$ is in a coherent superposition (linear combination) $\vert \psi \rangle = \sum_{i=1}^n a_i \vert \psi_i \rangle$ of (basis) elements $\vert \psi_i\rangle$ of that context.

The `measurement ancilla'---along with synonymous terms such as `provision', `component', or `arrangement'---should be represented by another state,
denoted as $\vert \varphi \rangle$.
This state, in relation to a suitable basis ${\vert \varphi_1 \rangle, \ldots, \vert \varphi_n \rangle}$,
can also be expressed as a coherent superposition: $\vert \varphi \rangle = \sum_{j=1}^n b_j \vert \varphi_j \rangle$.
When an interaction occurs between the `object' and the `measurement ancilla', the combined state $\vert \Psi \rangle$
\begin{equation}
\vert \Psi \rangle = \sum_{i,j=1}^n c_{ij} \vert \psi_i \rangle \otimes \vert \varphi_j \rangle =
\sum_{i,j=1}^n c_{ij} \vert \psi_i \varphi_j \rangle
\label{2024-u-vNsiqm}
\end{equation}
becomes a non-factorizable tensor product, meaning that the coefficients $c_{ij}$
cannot be written as products $a_i b_j$.

From now on, when referring to the `object' and the `measurement ancilla', apostrophes will be omitted. Since in entangled systems individuality is traded for relationality between individual components~\cite{zeil-99}, any conceptualization of a Heisenberg cut between these entangled constituents is a classical notion that may be maintained for all practical purposes (FAPP~\cite{bell-a}) but, strictly speaking, is not applicable.

\section{Infinite tensor products}

By recursively applying the von Neumann scheme for ideal quantum measurement~(\ref{2024-u-vNsiqm}), we can construct increasingly larger product spaces as more factors are added. To form a Hilbert space, we take the closure of this space under a suitable norm derived from the inner product. This process can be understood as taking the `double dual', or more specifically, the dual of the vector space of all bilinear forms on the vector spaces participating in the product~\cite{halmos-vs}.


\subsection{Elementary Tensors as Products of Basis Vectors}

Given a countable collection of Hilbert spaces
\(
\Big\{  \mathcal{H}_n \Big| \; 1 \le k_n \le d_n \in \mathbb{N} \Big\}
\),
let
\(
\Big\{ \vert k_n \rangle  \Big| \; 1 \le k_n \le d_n \in \mathbb{N} \Big\}
\)
be an orthonormal basis for each \(\mathcal{H}_n\),
where the dimension \(d_n\) of \(\mathcal{H}_n\) could be a finite positive integer or countably infinite.

An elementary tensor product \(\bigotimes_{n=1}^\infty \vert k_n \rangle \) is then given by
\[ \bigotimes_{n=1}^\infty \vert k_n \rangle
= \vert k_1 \rangle \otimes \vert k_2 \rangle \otimes \vert k_3 \rangle \otimes \cdots
= \vert k_1 k_2 k_3 \cdots \rangle
\]
where \(\vert k_n \rangle\) is the $k_n$th basis vector from \(\mathcal{H}_n\).


\subsection{Tensor Product Space}

Let $I$ be a countable (enumerable) infinite index set, identified with the set of all natural numbers, $\mathbb{N}$.
The labelling $n$ represents the $n$th subfactor of the tensor product.
For the sake of simplicity, from now on, we will consider the set of all natural numbers as our index set.
So, whenever we write, say, $\bigotimes_{n=1}^\infty | k_n \rangle $ we really mean $\bigotimes_{n \in I} | n \rangle$.

To form the tensor product space \(\bigotimes_{n=1}^\infty \mathcal{H}_n\)
\begin{itemize}

\item[(i)]
we start with elementary tensors \(\bigotimes_{n=1}^\infty \vert k_n \rangle \),
where $n$ labels the $n$th subfactor of the tensor product,
and $k_n$ represents the $k_n$th basis vector in \(\mathcal{H}_n\).

\item[(ii)]
We define the inner product on elementary tensors by the product of the individual inner products~\cite[Definition~II.5., p.~63]{Guenin1969}:
\[ \left\langle \bigotimes_{n=1}^\infty \vert k_n\rangle \middle| \bigotimes_{n=1}^\infty \vert j_n\rangle  \right\rangle
= \begin{cases}
    \prod_{n=1}^\infty \langle k_n \vert j_n \rangle_{\mathcal{H}_n} &  \text{converging,}\\
    0 & \text{otherwise.}
\end{cases}
\]
Nonvanishing inner products will later, in Subsection~\ref{2024-u-sectors}, serve as a criterion for vectors to belong to the same sector.

\item[(iii)]
We then consider finite linear combinations of these elementary tensors:
   \[ \sum_{i} c_i \left( \bigotimes_{n=1}^\infty \vert {k_n^{(i)}} \rangle \right), \]
   where \(c_i\) are complex coefficients and \(\vert k_n^{(i)} \rangle\)
are basis vectors in the elementary tensor product labeled by a countable (enumerable) index~$i$ (see discussion later).

\item[(iv)]
We finally obtain the Hilbert space \(\bigotimes_{n=1}^\infty \mathcal{H}_n\) by taking the completion of the space of finite linear combinations of elementary tensors with respect to the norm induced by the inner product, ensuring that the space is complete and satisfies the properties of a Hilbert space.

\end{itemize}

By defining elementary tensors as products of basis vectors from the bases of the factors,
a concrete and manageable set of elementary tensors is obtained, spanning the tensor product space.
This approach, derived from finite tensor products~\cite[Theorem~1, \S~24,25]{halmos-vs},
simplifies both the definition and the computation of the inner product, ensuring that the space \(\bigotimes_{n=1}^\infty \mathcal{H}_n\) has a well-defined Hilbert space structure.

However, this construction does not directly address the uncountable infinity of elementary products. To illustrate this,
we can draw an analogy with the representation of real numbers as expansions in an $n$-ary system, where they are encoded using a (finite) set of basis elements.
Just as Cantor's diagonal argument shows that the reals cannot be enumerated by any countable set of indices,
so too can we not enumerate the uncountable infinity of elementary products in the infinite tensor product.
In von Neumann's own words \textit{``generalisations of the direct product lead to higher
set-theoretical powers (G. Cantor's ``Alephs'')''}~\cite[S~4, p.~4]{vonNeumann1939}.
Following Von Neumann's `incomplete infinite direct products'~\cite[Chapter~4]{vonNeumann1939},
Thirring and Wehrl  define the infinite tensor product in terms of equivalence classes~\cite[\S~2]{thirring-1967}
(see also Thirring~\cite[Definition~II.4., p.~63]{Guenin1969}) discussed later in the context of sectorization.
One could even go so far as to suspect that many of the upcoming issues related to continuity originate from this fact.

\subsection{Violation of unitary equivalence}

In finite dimensions, unitarity is a property of a single operator, characterized by its ability to preserve the inner product and possessing an inverse equal to its conjugate transpose.
On the other hand, unitary equivalence is a relation between two operators or orthonormal bases, signifying that one can be transformed into the other via a unitary transformation.
Fundamentally, unitarity captures the properties of an individual operator, whereas unitary equivalence captures the relationship between two operators or orthonormal bases.

Infinite tensor products pose significant challenges to maintaining unitary equivalence, primarily due to difficulties in defining a consistent inner product,
achieving proper normalization, preserving the required topological structure of the Hilbert space, and managing unbounded operators.
These challenges make it problematic to uphold the fundamental principles of quantum mechanics,
including the ability to execute arbitrary unitary transformations within the Hilbert space.
Notably, certain dynamical processes, such as the interaction between sectors
(as explored in the infinite limit in subsection~\ref{2024-u-sectors}),
become severely restricted when confined to finite resources.

\subsubsection{Inner product and orthogonality}

The inner product is a crucial concept in quantum mechanics, ensuring that probability amplitudes are well-defined and that the evolution is unitary.
However, in an infinite tensor product space, defining an inner product that adheres to the properties of a Hilbert space poses significant challenges.
Key among these challenges are issues with convergence, which are closely tied to the orthogonality of states.

As long as the tensor product is finite, the inner product is well-behaved.
When we extend to an infinite tensor product, say
\( | \Psi \rangle = \bigotimes_{i=1}^{\infty}| \psi_i \rangle \) and \( |  \Phi \rangle = \bigotimes_{i=1}^{\infty} | \phi_i \rangle \),
the inner product would apparently be
\(
\langle \Psi \vert \Phi \rangle = \prod_{i=1}^{\infty} \langle \psi_i \vert \phi_i \rangle
\).

The central issue is whether this infinite product converges to a non-zero value or not.
Suppose, for the sake of demonstration, that each \( \langle \psi_i \vert \phi_i \rangle \)
is very slightly less than 1. As a consequence, the infinite product can converge to zero, and thus those
vectors which are only `slightly apart' appear orthogonal.
Formally, suppose that
\(
\langle \psi_i \vert \phi_i \rangle = 1 - \epsilon_i  = \delta_i
\),
or
\(
  \epsilon_i = 1 - \langle \psi_i \vert \phi_i \rangle
\),
where
\(
0 < \epsilon_i \ll 1
\).
For a large number of factors, the infinite product behaves approximately as
\begin{equation}
\langle \Psi \vert \Phi \rangle = \prod_{i=1}^{\infty} (1 - \epsilon_i) \approx \exp\left(-\sum_{i=1}^{\infty} \epsilon_i\right)
.
\label{2024-u-dwsp}
\end{equation}
If the series \( \sum_{i=1}^{\infty} \epsilon_i \) diverges (even if slowly), this product will converge to zero, that is,
\(
\prod_{i=1}^{\infty} (1 - \epsilon_i) \rightarrow 0.
\)


Furthermore, the inner product would also become zero for states $\vert \Psi \rangle$ and $\vert \Phi \rangle$
that differ in only a single or a finite number of the infinitely many subfactor components, where $\langle \psi_i \vert \phi_i \rangle = 0$, with all the rest being identical.
Additionally, there may be issues related to the phases, as explored by von Neumann~\cite{vonNeumann1939}
and by Van Den Bossche and Grangier~\cite{van-den-bossche-2023-a}.

This highlights the challenges infinite tensor products face in preserving a consistent inner product structure.
In numerous instances, the inner product may become undefined or yield counterintuitive outcomes,
contravening the anticipated properties of a Hilbert space and subsequently impacting unitary equivalence.

\subsubsection{Norm}

Issues with inner products in turn translate into problems with normalization, as
the polarization identity
expresses the inner product of two vectors in terms of the norm of their differences; that is,
\(
\langle  \Psi \vert  \Phi \rangle
=
\frac{1}{4}\left[
\|   \Psi +  \Phi  \|^2
-
\|   \Psi -  \Phi  \|^2
+ i
\left(
 \|   \Psi - i \Phi  \|^2
-
\|   \Psi + i \Phi  \|^2
\right)
\right]
\).
Thus, for $\langle  \Psi \vert  \Phi \rangle =0$,
$\|   \Psi +  \Phi  \|^2 = \|   \Psi -  \Phi  \|^2$,
and
$\|   \Psi - i \Phi  \|^2
-
\|   \Psi + i \Phi  \|^2  $.
This is true for finite tensor products but not necessarily for infinite ones
if, as before,
vectors \(  \vert \Psi \rangle \) and \(  \vert \Phi \rangle \)
represent physically distinct states located `close to each other',
such that the subfactors $\langle \psi_i \vert \phi_i \rangle = 1 - \epsilon_i$  where
\(
0 < \epsilon_i \ll 1
\).

As before this applies also to states $\vert \Psi \rangle$ and $\vert \Phi \rangle$
differing in only a single one or a finite number of infinitely many subfactor components
where $\langle \psi_i \vert \phi_i \rangle = 0$, with all others remaining identical.



\subsubsection{Bounded Operators}

Let us consider an example involving an infinite tensor product of projection operators to illustrate issues
with bounded operators on infinite tensor products.

Consider the Hilbert space \(\mathcal{H} = \mathbb{C}^2\) (2-dimensional complex space).
Let \(E\) be the rank-one projection operator onto the subspace spanned by the vector
\(\vert \uparrow \rangle = \begin{pmatrix}1 , 0\end{pmatrix}^\intercal\), with
\( E = \text{diag}\begin{pmatrix} 1 , 0 \end{pmatrix}\).
Now, consider the operator \(F = \bigotimes_{n=1}^{\infty} E\), which is the infinite tensor product of \(E\) with itself.

Initially, it may seem that \(E\) being a projection operator with \(\|E\| = 1\), \(F\) would be a well-defined bounded operator with \(\|F\| = 1\). However, this is not the case. To understand why,
let us examine the action of \(F\) on specific vectors.

Let us represent a general vector in the infinite tensor product space as \(\vert \psi \rangle = \bigotimes_{n=1}^{\infty} \vert \psi_n \rangle\), where \(\vert \psi_n \rangle\) are vectors in \(\mathcal{H}_n\).
For simplicity, assume each \(\vert \psi_n \rangle\) is a normalized vector in \(\mathbb{C}^2\).

When \(F\) is applied to \(\vert \psi \rangle\), we get
\(
F \vert \psi \rangle = \bigotimes_{n=1}^{\infty} E \vert \psi_n \rangle
\).

Since \(E\) projects onto \(\begin{pmatrix} 1 , 0 \end{pmatrix}^\intercal\), the resulting vector will be non-zero only if each \(\vert \psi_n \rangle\) has a component along \(\begin{pmatrix} 1 , 0 \end{pmatrix}^\intercal\). In an infinite product, the probability of each \(\vert \psi_n \rangle\) having a non-zero component along \(\begin{pmatrix} 1 , 0 \end{pmatrix}^\intercal\) diminishes rapidly, effectively leading to the result that \(F \vert \psi \rangle = 0\) for almost all \(\vert \psi \rangle\).

For instance, consider the vector \(\vert \psi \rangle = \vert \uparrow \rangle \otimes \vert \uparrow \rangle \otimes \cdots\), where
\(F \vert \psi \rangle = \vert \psi \rangle\) and \(\| F \vert \psi \rangle \| = \|\vert \psi \rangle\| = 1\).
On the other hand, for any vector
containing a component orthogonal to \( \vert \uparrow \rangle = \begin{pmatrix} 1 , 0 \end{pmatrix}^\intercal \),
such as the spin-down state $ \vert \downarrow \rangle = \begin{pmatrix}  0,1  \end{pmatrix}^\intercal $, in at least one factor,
\(F\) maps it to the zero vector. For example, for the vector
\(
\vert \varphi \rangle = \vert \uparrow \rangle \otimes \vert \uparrow \rangle \otimes \cdots \otimes
\vert \downarrow \rangle \otimes \vert \uparrow \rangle \otimes \cdots,
\)
we have \(F \vert \varphi \rangle = 0\).

This demonstrates that the infinite tensor product \(F = \bigotimes_{n=1}^\infty E\) does not behave as a well-defined bounded operator in the infinite tensor product space. Although \(F\) leaves certain vectors unchanged---those entirely within the span of \(\begin{pmatrix} 1 , 0 \end{pmatrix}^\intercal\)---it maps any vector with even a single orthogonal component to zero.
This behavior leads to some counterintuitive physical properties because \(F\) is extremely sensitive to changes in its input: changing even one factor from \(\begin{pmatrix}1 , 0\end{pmatrix}^\intercal\) to any other vector results in mapping the vector to zero.

Furthermore, the behavior of \(F\) is consistent with finite tensor products of \(E\). In both finite and infinite cases, the result is a rank-one projection. However, the key difference is that in the infinite case, this leads to a projection onto a one-dimensional subspace of an infinite-dimensional space, which has some unique properties.

\subsection{Sectorization}
\label{2024-u-sectors}


Von Neumann's concept of `incomplete infinite direct products'~\cite[Chapter~4]{vonNeumann1939}, as reflected in the notion of superselection sectors in algebraic quantum field theory~\cite{haag-1964,doplicher-1969,doplicher-1971,doplicher-1974,haag-1996}, provides a solution to the problem that a single (or finitely many) subfactor(s) could nullify the inner product by `grouping' vectors that differ from each other in only finitely many subfactors, or are otherwise `close to' each other.
These groupings are mutually orthogonal and can be demonstrated to be equivalence classes referred to as sectors.
This implies that vectors from different sectors differ in an infinite number of subfactors and are orthogonal in the sense that their scalar product is zero.
In von Neumann's own words,
\textit{``What happens could be described in the quantum-mechanical
terminology as a `splitting up' of [[the full tensor product]] into `non-intercombining
systems of states', corresponding to the `incomplete' direct products''~\cite[\S~6, p.~4]{vonNeumann1939}}

Formally, within each sector are only those infinite tensor products that are located `close to' each other,
such that their deviations from each other are `small'. Two vectors $ \vert \Psi \rangle $ and $ \vert \Phi \rangle $
are in the same sector if they are equivalent, denoted by $ \vert \Psi \rangle \sim \vert \Phi \rangle $,
when all but finitely many subfactors are either equal or unitary equivalent and to or
`close to' one another (only a unitary transformation apart); that is,
with the notation from Eq.~(\ref{2024-u-dwsp}) we require~\cite[Eq.~(9)]{van-den-bossche-2023-a}
\begin{equation}
\sum_{i=1}^{\infty} \left( 1 - \langle \psi_i \vert \phi_i \rangle \right)
=\sum_{i=1}^{\infty} \epsilon_i
\le
\sum_{i=1}^{\infty} \left| 1 - \langle \psi_i \vert \phi_i \rangle \right|
< \infty
.
\end{equation}
(With $0< \epsilon_i \ll 1$ as above this would be an equality.)
This condition ensures that the product
$ \prod_{i=1}^{\infty} \langle \psi_i \vert \phi_i \rangle $
converges to a non-zero value within the same sector.
The inner product of infinite tensor products
belonging to different sectors vanishes.

For finite tensor products resulting in finite-dimensional Hilbert spaces,
sectorization has no meaningful relevance: Since, in finite dimensions, all orthonormal bases are unitarily equivalent.


Therefore, for infinite tensor products, instead of directly dealing with the entire infinite tensor product space,
one should consider regions or sectors within it. These sectors are equivalence classes of vectors that differ only by a finite number of components
in the tensor product, or are otherwise close to (unitary equivalent) each other. In this framework, it is postulated that these mathematically distinct sectors correspond physically to different `global' or `macroscopic'
configurations of the system, such as pointer states of a measurement apparatus~\cite{hepp-1972}.

States in different sectors cannot be coherently superposed by finite (unitary) means.
One may say that, with respect to these finite unitary means,  `coherence is lost'.
Hepp even went so far as to state that \textit{``leads to macroscopically  different `pointer positions' and to a rigorous
'reduction of the wave packet'~''}~\cite{hepp-1972}.
Let us demonstrate this with an example.
Consider an infinite array of spin-$\frac{1}{2}$ particles,
where each particle has a Hilbert space $\mathcal{H} = \mathbb{C}^2$,
spanned by the states $\vert \uparrow\rangle$ (spin up) and $\vert \downarrow\rangle$ (spin down).
The entire system is then described by the infinite tensor product of these spaces
\(
\mathcal{H}_{\text{total}} = \bigotimes_{i=1}^{\infty} \mathbb{C}^2_i
\).

In such a setup, a  {sector} can be defined by specifying the `macroscopic' behavior of the system, such as the average magnetization:

\begin{itemize}
    \item  {Sector A (All Spins Up)}: Consider the state where every spin is up; that is,
    \(
    \vert \psi_{\text{up}}\rangle = \vert \uparrow\rangle \otimes \vert \uparrow\rangle \otimes \vert \uparrow\rangle \otimes \cdots
    \).
    This state belongs to a sector where all spins are aligned up.

    \item  {Sector B (All Spins Down)}: Similarly, consider the state
    \(
    \vert \psi_{\text{down}}\rangle = \vert \downarrow\rangle \otimes \vert \downarrow\rangle \otimes \vert \downarrow\rangle \otimes \cdots
    \)
    where every spin is down.
    This state belongs to a different sector where all spins are aligned down.

    \item  {Sector C (Mixed Alignment)}: Now consider a state where half the spins are up and half are down, such as
    \(
    \vert \psi_{\text{mixed}}\rangle = \vert \uparrow\rangle \otimes \vert \downarrow\rangle \otimes \vert \uparrow\rangle \otimes \vert \downarrow\rangle \otimes \cdots
    \).
    This state belongs to yet another sector, where the system exhibits a different macroscopic behavior.
\end{itemize}

To illustrate the challenges encountered, let us attempt to superpose states from different sectors.

\begin{itemize}
    \item  {Within the Same Sector}:
    Superpositions of states within the same sector are possible. For example, superpositions of states that differ by a finite number of spins can be physically meaningful, such as
        \(
        \vert \psi\rangle = \alpha \vert \uparrow\rangle \otimes \vert \uparrow\rangle \otimes \vert \uparrow\rangle \otimes \cdots + \beta \vert \uparrow\rangle \otimes \vert \downarrow\rangle \otimes \vert \uparrow\rangle \otimes \cdots
        \)
        Both states essentially belong to the same `all spins up' sector with minor fluctuations.

    \item  {Across Different Sectors}:
     Attempting to superpose states from different sectors, such as:
        \(
        \vert \phi\rangle = \alpha \vert \psi_{\text{up}}\rangle + \beta \vert \psi_{\text{down}}\rangle,
        \)
        results in a superposition that is  {not physically meaningful}.
States from different sectors (like all spins up versus all spins down) represent distinct macroscopic configurations,
and there is no way to coherently combine them in an infinite system using finite means.

\end{itemize}

Let us now address the question of why coherence---the ability to linearly superpose states from different sectors---is lost
in an infinite tensor product space.
It is essential to note that different sectors are orthogonal: States from different sectors (like all up versus all down)
become orthogonal in the limit of an infinite number of particles.
This orthogonality is a reflection of the fact that they represent fundamentally distinct physical configurations.
Furthermore, there exists no observable capable of coherently mixing states from different sectors,
implying that any attempt to superpose them would not result in interference effects.
In the limit case of infinite tensor product states, the system effectively `forgets' any phase relationship between these states,
leading to a loss of coherence.

In the infinite tensor product of spin-$\frac{1}{2}$ systems,
sectors correspond to different macroscopic configurations of spins---for instance, all up, all down, and mixed.
States from different sectors cannot be coherently superposed
because they are orthogonal and no (finite) transformation connects them---they tend
to `crystallize' or `decohere' into different macroscopic domains or realms---leading to a loss of coherence. This example illustrates how sectors naturally arise in infinite tensor products and
why superpositions across sectors are not physically meaningful.
With this kind of sectorization, or transition into different sectors, global unitarity with respect to finite unitary means is lost.

\subsection{Factorization}
\label{2024-u-factors}

Von Neumann algebras, also known as \( W^*\)-algebras,
are operator algebras that are classified into types I, II, and III,
introduced by von Neumann and Murray~\cite{Murray1936}.
These algebras are closed under addition, operator and scalar multiplication, and contain the identity.
The `star' symbol $^*$ indicates closure under adjoint transformations.
They are also closed in the weak operator topology with respect to operator sequences
converging towards a limit, ensuring that they are complete under this topology.

A von Neumann algebra \( \mathcal{M} \) is called a \textit{factor} if its center \( \mathcal{Z}(\mathcal{M}) \)---the set of all operators in \( \mathcal{M} \) that commute with every operator in \( \mathcal{M} \)---consists only of scalar multiples of the identity operator.
Factors are indecomposable in the sense that they cannot be decomposed into a direct sum of two non-trivial von Neumann algebras.
Furthermore, any von Neumann algebra can be written as a direct sum of its factors~\cite{Penington2022Dec}.


Von Neumann factors are classified into three types: I, II, and III. This classification is based on the structure of projections in the algebra and the trace properties.


Type I factors are those that are isomorphic to all bounded operators on a Hilbert space:
\begin{itemize}

\item {Type I\(_n\):} The factor is isomorphic to \( M_n(\mathbb{C}) \), the algebra of \( n \times n \) matrices over the complex numbers. These factors correspond to finite-dimensional Hilbert spaces.

\item {Type I\(_\infty\):} The factor is isomorphic to \( \mathcal{B}(\mathcal{H}) \),
the algebra of all bounded operators on an infinite-dimensional separable Hilbert space \( \mathcal{H} \).
These factors act on Hilbert spaces with countably infinite dimension.

\end{itemize}

It is reasonable to identify (orthogonal) projections of type I$_n$
factors with elements of orthonormal bases, or equivalently, with contexts, blocks in quantum logic, or maximal operators.
This identification is supported by~\cite[Satz~8]{v-neumann-31} (see also~\cite[\S~82]{halmos-vs}).
Type I\(_n\) factors  are the only ones in finite dimensional Hilbert space.
They have minimal orthogonal projections (self-adjoint and idempotent) that correspond
to one-dimensional subspaces of the Hilbert space,
as well as convex combinations thereof (projecting into higher-dimensional subspaces).
Indeed, any sequence of mutually orthogonal (orthogonal) projections $|\psi_i\rangle$,
combined with any sequence of probabilities $p_i \in [0,1]$ satisfying $\sum_{i=1}^{k} p_i = 1$ where $k \leq n$,
forms a density operator $\rho = \sum_{i=1}^{k} p_i |\psi_i\rangle\langle\psi_i|$.

In terms of quantum mechanical states, this amounts to both pure and mixed states~\cite{sorce-2023}.
Note that, in the context of a finite-dimensional Hilbert space, the trace of a
$k$-dimensional projection in an $n$-dimensional space (where $k\le n$) is simply the positive integer $k$.

In terms of entanglement, type I\(_n\) factors can (but need not) represent a finite number of entangled particles.
Type I\(_\infty\) factors are infinite-dimensional.


Type II factors are characterized by their occurrence in infinite-dimensional Hilbert spaces.
In contrast to type I factors, they are considered `diffuse', meaning they lack minimal projections,
which are projections onto one-dimensional subspaces or convex combinations thereof~\cite{houdayer-aittIIf}.
This property is often linked to mixed states in the context of quantum mechanical states~\cite{sorce-2023}.

\begin{itemize}

\item {Hyperfinite type II\(_1\) factor:}
In contrast to type I factors, the entanglement in a hyperfinite type II\(_1\) factor is more `diffuse', making it impossible to identify individual entangled states.
For example, the factor might comprise an infinite number of qubit pairs, with all but a finite number of pairs in a maximally entangled state~\cite{Penington2022Dec}.

\item {Type II\(_\infty\) factor:} This factor is simply the tensor product of a type II\(_1\) factor and a type I\(_\infty\) factor.

\end{itemize}

Despite being diffuse, the trace of a projection in a type II factor is still faithful, normal, and semi-finite.
A faithful trace is one that does not vanish on any non-zero positive element of the von Neumann algebra.
A normal trace respects the convergence of operators (in the weak topology), ensuring that the trace of a limit of operators equals the limit of their traces.
A semi-finite trace is one such that for any positive element, there is a non-zero `sub-element' on which the trace is finite.
For type II\(_1\) factors, the trace assigns a value in the continuous interval \([0,1]\), where 0 corresponds to the zero projection and 1 corresponds to the identity projection.
This trace function behaves like a measure of `dimension' but is not tied to integer dimensions as in finite-dimensional spaces.
For type II\(_\infty\) factors, the trace can take values in \([0,\infty]\).

%
%

Type III factors are characterized by the absence of faithful normal semi-finite traces.
In the context of quantum mechanical states, this implies that states on Type III factors
cannot be represented by density operators in the conventional sense.
The distinction between pure and mixed states becomes more intricate,
as all normal states on a Type III factor are, in some sense, `mixed.'
Nevertheless, notions of pure states (as extreme points of the state space) and mixed states still exist,
but they exhibit different behaviors compared to those in Type I or II factors.
In terms of entanglement, we can expect `infinite entanglement' but also `infinite fluctuations' in this entanglement~\cite{Penington2022Dec}.



The origin of the term `factor' may come from a tensor product factorization:
Suppose
$\mathcal{H} = \mathcal{H}_1 \otimes \mathcal{H}_2$.
Then
$F_1 = \mathcal{B}(\mathcal{H}_1) \otimes \mathbb{1}$
and
$F_2 = \mathbb{1} \otimes \mathcal{B}(\mathcal{H}_2)$
are factors of
$\mathcal{B}(\mathcal{H})$~\cite[Exercise~3.3.8]{Jones-vNA}.

Unitary equivalence preserves the type of the algebra and does not change it.
As a consequence, there are no unitary operators, permutations, or any other operations
within the framework of von Neumann algebras that can convert, say, a type I factor into a type II or III factor.

To facilitate transitions between factor types, more powerful tools than unitary operations are needed.
One such tool (among others) is the use of inductive limits, which enable the construction of large and complex algebras from simpler, smaller ones~\cite{houdayer-2009}.
To transition from type I factors to type II factors, sequences or non-trace-preserving embeddings that 'diffuse' the trace structure are required.

\section{Nested Wigner's friends as infinite tensor products}

Nesting or chaining refers to the repeated and iterated application of the von Neumann type measurement-by-entanglement,
as formalized by Equation~(\ref{2024-u-vNsiqm}), as expressed to first order by Wigner~\cite{wigner:mb}.
As a consequence, we end up with large and potentially infinite tensor products.
It is tempting to ascribe this measurement conceptualization to von Neumann~\cite{Taub:1961:JNCc,vonNeumannCompendium}.

Grangier and Van Den Bossche have recently proposed that the apparent loss of coherence
in such situations is attributable to sectorization
and the consequent loss of unitary equivalence within finite systems~\cite{Grangier-2020,van-den-bossche-2023-a,van-den-bossche-2023-b,van-den-bossche-2023-c},
as previously discussed in subsection~\ref{2024-u-sectors}.
According to their proposal, sectorization is a physical process in infinite algebras where separable
sectors correspond to `classical outcomes' and `macroscopic states' of pointers~\cite{hepp-1972,bub-1988,bub-2015}.
The ``context'' is physically realized by the choice of a measurement basis, which is determined by the experimental arrangement.
An ``observation'' then corresponds to an interaction that entangles the system with the measurement apparatus, projecting the state into this chosen basis.

While a single such interaction is a standard unitary evolution of the combined system, the formalism of infinite tensor products suggests that an infinite sequence
of such interactions forces the total state into one of the newly formed sectors.
 This transition to a different, orthogonal sector is what is meant by a ``shift'' that establishes a new macroscopic outcome.
A subsequent measurement in an incompatible basis (a new context) would similarly drive the system into yet another distinct sector,
a process described here as `reshuffling' or `scrambling' of contexts.

The connection between sectors and factors remains an open question.
A fundamental difference is that factors pertain to algebras of operators---specifically,
(generalized) density operators if a trace exists---while sectors pertain to (unitarily equivalent) elements or subspaces of Hilbert space.

For type I factors, some of these operators can be interpreted as pure
(that is, `minimal' one-dimensional orthogonal projection operators) or mixed states.
Indeed, intuition from finite tensor products suggests that, through spectral decomposition of the operators in a factor,
the respective orthogonal projections in type I$_n$ factors correspond to elements of contexts
(associated with maximal operators, see~\cite[Satz~8]{v-neumann-31}).
This implies that they refer to pure states spanned by vectors in the respective Hilbert space.

In finite dimensions, sectors do not carry much significance, as all vectors are unitarily equivalent.
However, for infinite tensor products, sectorizations `form naturally' through the unitary equivalence of vectors
(or their span, and the associated `minimal' one-dimensional orthogonal projection operators)
and can be associated with macroscopic quantities.

The transition between different sectors does not correspond to any unitary transformation, as these sectors are not unitarily equivalent.
This results in an apparent loss of coherence, as different sectors cannot be in coherent superposition.
Measurements with mismatched pre- and post-selection are linked to distinct sectorizations of the Hilbert space.
In the context of infinite tensor products, such `context translations' cannot be achieved through unitary operations.

Factorization offers another potential mechanism beyond unitary evolution:
The infinite limit of nested Wigner's friends, facilitated by entanglement, could enable transitions between distinct factor types.
Consequently, both factorization and sectorization, in the infinite limit, might contribute to a loss of unitary equivalence and decoherence.

Furthermore,
as previously discussed, any nesting construction is highly susceptible to
alterations in the focus of observations by Wigner's friends---specifically, with regard to changes in the sequence of entangled basis vectors.
This sensitivity arises not only from potential state changes within a (Type I) factor,
but also from the mismatches and entanglements that occur between infinite sequences of nested
von Neumann measurements,
which can lead to transitions into distinct sectors and factors.
Consequently, even the slightest mismatch and change in nested observables cumulatively leads to a complete loss of information
about the initial state (preparation).

More explicitly, as has been pointed out earlier in the context of difficulties in defining the inner product, any slight mismatch between (successive) friends' measurements `builds up' into a total loss of coherence.
This results in a vanishing inner product \(\langle \Psi \vert  \Psi' \rangle\) which converges to zero, indicating that the product states
\(\vert \Psi\rangle\) and \(\vert \Psi'\rangle\)
are orthogonal even if each single mismatch characterized by \(\langle \psi_i \vert  \psi_i' \rangle\) is very close to 1.
This type of `decoherence' is gradual and smooth in the sense that there is no abrupt discontinuous
transition---indicating a well-defined, localizable Heisenberg cut at some scale---but a gradual, continuous loss
of information about the initial state: Let $0 \ll \vert  \langle \psi_i \vert  \psi_i' \rangle \vert  = \delta_i < 1$ be this match per friends $i$ and $i'$, then
\begin{equation}
\vert  \langle \Psi \vert  \Psi' \rangle \vert
= \prod_{i=1}^{\infty} \vert  \langle \psi_i \vert  \psi_i' \rangle \vert
= \prod_{i=1}^{\infty} \delta_i
=0.
\end{equation}
One could also interpret $\epsilon_i$ in  $\delta_i= 1-\epsilon_i$ as a (measure of) stochastic `input' per Wigner's friend $i$ that contributes to a context translation~\cite{svozil-2003-garda,svozil-2013-omelette} but introduces additional input from the Wigner's friend (environment). This is particularly true if Wigner's friends attempt to `measure'
a state in which the quantum system is not prepared~\cite{zeil-99}.

This model diverges from the reduction model of Hepp~\cite{hepp-1972,bub-2015} and the recent papers by
Grangier and Van Den Bossche~\cite{Grangier-2020,van-den-bossche-2023-a,van-den-bossche-2023-b,van-den-bossche-2023-c}
in that it proposes a sequence of mismatch measurements by Wigner's friends that ultimately transcends sectors or even factors,
and does not depend on sectorization, that is, the creation of sectors interpretable as macroscopic `pointers'.

Bell's argument~\cite{Bell-1975} against transfinite recursion remains valid for an infinite number of Wigner's friends.
However, his later FAPP argument~\cite{bell-a}---that, although any Heisenberg cut is relative, it exists for all practical purposes
and experimental capacities---can be maintained.
I concur that any finite number of Wigner's friends does not lead to a violation of unitary equivalence,
and thus state reduction or decoherence.


Grangier and Van Den Bossche circumvent Bell's argument ontologically by positing that a dual quantum and classical description is necessary to understand quantum mechanics~\cite{van-den-bossche-2023-c}.
They argue that the mathematical formalism should provide a consistent description, rather than a complete (isomorphic) representation of reality.
In this context (see also the quote by Hertz mentioned later),
the use of mathematical infinities becomes a valid tool for description.

Another approach to addressing Bell's argument involves the concept of infinity processes, as discussed by Weyl~\cite[pp.41,42]{weyl:49}
in the context of Zeno's paradoxes (of infinite divisibility). If we consider the continuum
(or at least the infinite divisibility of space and time), \textit{``if the segment of length 1 really consists of infinitely
many subsegments of lengths $1/2, 1/4,1/8, \ldots$, as of `chopped-off'
wholes, then it is incompatible with the character of the infinite as the
`incompletable' that Achilles should have been able to traverse them all.''}
This implies that even for classical motion in a continuum to be possible, we require transfinite capacities.
This concept can be applied to the infinite nesting of Wigner's friends by considering this nesting or chaining
as the effective oneness which we experience; resulting in irreversible measurements in the transfinite limit.



\section{Historical Analogues}

To motivate the use of infinite limits as a tool for explaining emergent irreversibility, this section presents analogies from other fields of mathematics and physics where infinite processes lead to qualitatively new phenomena not present in their finite counterparts. These examples are intended as heuristic support to demonstrate a recurring structural parallel, rather than as direct physical evidence for the quantum mechanical argument.
This section explores several related but distinct concepts that have been investigated in various areas of physics.

\subsection{Infinity and Transfinite Capacities}

This concept is similar to the convergence of sequences of rational numbers to an irrational number in the real numbers.
For instance, consider the continued fraction or the binomial series expansions
\(
\sqrt{2} = (1 + 1)^{1/2} = \sum_{n=0}^{\infty} \binom{1/2}{n} \cdot 1^n = 1 + \frac{1}{2} \cdot 1 - \frac{1}{8} \cdot 1^2 + \frac{1}{16} \cdot 1^3 - \cdots
\)
of \(\sqrt{2}\), truncated at various points.

Another analogue is from recursive analysis: Specker sequences of computable numbers converge to an uncomputable limit~\cite{specker49,specker-ges,kreisel,simonsen-2005}. One example is Chaitin's constant, the halting probability of prefix-free program codes on a universal computer~\cite{rtx100200236p,calude-dinneen06}, whose rate of convergence is tied to the halting time, and therefore, `grows faster' than any computable function.

Many of these metamathematical results are based on Cantor's diagonal argument~\cite{book:486992}, which demonstrates that, `in the limit, enumerable sets become non-enumerable continua'.

\subsection{Statistical physics}

Loschmidt's \textit{Umkehreinwand}~\cite{darrigol-2021} poses a challenge to the concept of irreversible processes at the macroscopic level, given the time-reversibility of microphysical laws. Loschmidt argued that if the microscopic laws are reversible, then any macroscopic process should also be reversible if we could precisely reverse the velocities of all particles in a system. This appears to contradict our everyday experience of irreversible processes and the postulate of the increase of entropy.

The canonical response to the \textit{Umkehreinwand} may seem evasive: while technically correct, due to statistical-probabilistic considerations, the \textit{Umkehreinwand} is means-relative~\cite{Myrvold2011237} and therefore only FAPP~\cite{bell-a} invalid. This is exemplified in Maxwell's pragmatic approach, avoiding detailed inquiries about individual molecules that would complicate the argument~\cite{Maxwell-1879,garber}: \textit{``avoiding all personal inquiries [[about individual molecules]] which would only get me into trouble.''}

One example of `irreversibility-in-the-limit' is the computation of $\sqrt{2}$, as mentioned in the aforementioned two examples:
the continued fraction expansion yields
$1,\frac{3}{2},\frac{7}{5},\frac{17}{12},\frac{41}{29},\frac{99}{70},\frac{239}{169},\frac{577}{408},\frac{1393}{985},\frac{3363}{2378},\ldots , \sqrt{2}$,
whereas the binomial series expansion yields
$1,\frac{3}{2},\frac{11}{8},\frac{23}{16},\frac{179}{128},\frac{365}{256},\frac{1439}{1024},\frac{2911}{2048},\frac{46147}{32768},\frac{93009}{65536}, \ldots, \sqrt{2}$.
Suppose that we delete all common terms from the two series. Then we end up with two series that are different, yet their limit is the same.
(Alternatively, take just the binomial series and rescale its summands by adding the term $1/n$ to each summand.)
Once the limit is reached, and no memory is maintained, it is impossible to determine which of the two series the result originates from~\cite{Svozil-2023-axioms12010072}.

\section{Summary}

We have presented both formal and pragmatic (FAPP) arguments for converting unitary, reversible von Neumann-Everett type 2 processes into non-unitary, irreversible type 1 processes. This conversion utilizes infinite tensor products, which, unlike in finite-dimensional Hilbert spaces, are not bound by unitarity. While objections may arise regarding the operational correspondence of infinite mathematical processes~\cite{bridgman}, a more practical approach involves considering finite subsequences or prefixes of these constructions.

These prefixes can be viewed as nested, iterated, or chained Wigner's friends, each encountering growing challenges in retrieving the original information from a quantum state, particularly when there is a discrepancy between state preparation and measurement. This phenomenon is reminiscent of environmental monitoring, resulting in quantum decoherence~\cite{schlosshauer-2005,schlosshauer-2019}, where environmental interactions lead to a loss of quantum coherence.

Parallels can also be drawn to noise introduction in micro-state amplification~\cite{Glauber-cat-86}, which illustrates the quantum no-cloning theorem and the disruption of quantum states through interactions or measurements.

The proposed mechanism can be contrasted with other solutions to the measurement problem. Unlike objective collapse models, it does not postulate a new physical collapse mechanism. In contrast to standard Everettian interpretations, the sector structure implies that not all branches of the universal state are equivalent, as transitions between them are forbidden by finite unitary means. And while the model shares features with decoherence, which also relies on interaction with a large environment, the use of infinite tensor products provides a formal basis for a strict, rather than merely practical (FAPP), breakdown of unitary equivalence between macroscopic outcomes.

The framework presented here also shares significant conceptual ground with standard decoherence theory, yet it is distinguished by a crucial formal difference.
Both approaches attribute the loss of quantum coherence to the system's entanglement with a larger, more complex entity---be it a physical environment or, in this paper's model, an infinite chain of observers.
The selection of a preferred basis (the 'pointer basis' in decoherence) is analogous to the formation of sectors corresponding to macroscopic outcomes.
 The primary divergence, however, lies in the nature of the resulting irreversibility. Standard decoherence describes a practical (FAPP) process within a finite, albeit large, system-environment composite.
The global evolution remains unitary, and coherence is merely 'leaked' into the environmental degrees of freedom,
 becoming locally inaccessible but never truly destroyed.
In contrast, the mechanism proposed here, leveraging the mathematical properties of the infinite tensor product limit, describes a formal and absolute breakdown of unitary equivalence.
The emergence of sectors is not a matter of information being difficult to retrieve;
it is a structural feature of the infinite-dimensional Hilbert space where states in different sectors are mathematically non-interconvertible by finite unitary means.
Thus, while decoherence explains the \emph{appearance} of classicality in a fundamentally unitary world, this paper's formalism offers a mathematical route to the \emph{emergence} of genuine, non-unitary irreversibility in the thermodynamic limit.

Schr\"odinger's `jellification' argument~\cite{schroedinger-interpretation} emphasizes the possibility of
unobserved quantum states `spreading' as coherent superpositions without being fixed by irreversible measurement.
Nesting Wigner's friends provides a solution characterized by three key aspects:
(i) The incorporation of environmental information, unrelated to the original state, leads to FAPP uncontrollable
(but not irreducible~\cite{zeil-05_nature_ofQuantum}) systematic stochasticity,
successfully converting the original (preselected) state into the measured (postselected) state.
(ii) Sectorization, which involves the effective orthogonalization and partitioning of the Hilbert space
into macroscopic regions corresponding to measurement outcomes, illustrates the practical difficulties of maintaining coherence when scaling up
the system to macroscopic dimensions~\cite{Grangier-2020,van-den-bossche-2023-a,van-den-bossche-2023-b,van-den-bossche-2023-c}.
(iii) Factorization, occurring according to the depth and modes of entanglement among the Wigner's friends,
additionally contributes to the emergence of classical-like behavior.

The processes of sectorization, which leads to the emergence of macroscopic observables, and factorization, which involves entanglement and the perception of isolated measurement outcomes,
both contribute to the loss of coherence and the formation of non-unitarily equivalent states in the infinite tensor product limit.
Given the pivotal role of entanglement in both nested Wigner's friends and factorization,
it is not totally unreasonable to conjecture a connection between these two phenomena.

Resolutions of the quantum measurement problem and the \textit{Umkehreinwand} in statistical physics through means relativity entail significant epistemological commitments. Previous attempts to simulate measurement processes using von Neumann algebras, such as those by Hepp~\cite{hepp-1972}, have faced criticism for relying on transfinite concepts without operational validation~\cite{Bell-1975,bub-2015}. Nonetheless, these findings could be reconciled by adopting the perspective that ``(FAPP) Infinity (FAPP) Does It.''

Classical analysis, recursive function theory, and von Neumann algebras offer potential ontological frameworks or 'escape routes'
from uniform reversibility and unitarity. However, their viability depends on the acceptance of
infinite limits as meaningful physical concepts~\cite{wigner}.
Modern resolutions of Zeno's and the Eleatics' paradoxes suggest that
without infinite limits and transfinite capacities, there is no motion in a continuum.

Another straightforward pragmatic approach could be considered:
Since this paper employs infinite processes to dispel unitary equivalence,
one could avoid the infinite limit by transcribing the discussion into the framework of `for all practical purposes' (FAPP).
For operationalists who prefer to avoid the use of strict limits,
the term `limit' can be substituted with `FAPP unboundedness' or `too-large-to-handle,'
and the symbol $\infty$ can be replaced with $\infty_\text{\tiny FAPP}$.

Yet, we must remain cognizant that both FAPP and transfinite irreversibility remain mathematical constructs,
Hertz's `images of our imagination'~\cite{hertz-94e} which ultimately
are justified by their practical usefulness and correspondence with phenomenology.
We should therefore exercise caution, not conflating the practical utility of our models with absolute certainty about physical reality.

Quantum erasure arguments~\cite{PhysRevA.25.2208,greenberger2,kim-2000,Ma22012013}
and the Humpty-Dumpty problem~\cite{engrt-sg-I,engrt-sg-II,Englert2013} further illustrate the challenges of reversibility and state reconstruction.
In a similar manner to classical statistical arguments, a macrostate corresponds to numerous microstates, making reversal attempts futile and mirroring the quantum context.

In my opinion, we cannot accept classical irreversibility without accepting irreversible quantum measurement; conversely,
FAPP insistence on classical and quantum reversibility expresses the same resistance towards transfinite, possibly nonconstructive means.
Pointedly stated, the central question in this comparative aspect becomes:
What is a viable position towards Loschmidt's \textit{Umkehreinwand},
and how does this stance translate to quantum measurement?
Whatever answer one might feel comfortable with regarding classical irreversibility,
one may apply its analog to quantum measurement irreversibility.
In both domains, the conceptualization~\cite{anderson:73} of macroscopic observables through grouping and sectorization,
as well as the entanglement-driven factorization in the infinite limit, challenge our notions of reversibility.

Ultimately, the parallel between classical and quantum irreversibility underscores a fundamental unity in physics,
while also highlighting the epistemological challenges and commitments we face in bridging the gap between our
finite experimental capabilities and the infinite limits of our theoretical constructs.

\begin{acknowledgments}
I would like to express my gratitude to Philippe Grangier and Mathias Van Den Bossche for their patient explanations of certain concepts in earlier drafts. Any remaining errors are, however, solely due to my own limitations.

I had the opportunity to meet Eugene Wigner personally during the 1st International Seminar on Nuclear War, held in Erice, Italy, from 14 to 19 August 1981.
However, I did not discuss his ``remarks on the mind-body question''~\cite{wigner:mb} with him.

The author declares no conflict of interest.

The AI assistants ChatGPT from OpenAI, mistral chat from mistral.ai, and Claude from Anthropic (partly through VSC enhanced Continue and Cursor) were used for formulating parts of the argument,
symbolic transformations into \LaTeX, and grammar and syntax checks.

This research was funded in whole or in part by the Austrian Science Fund (FWF) Grant DOI: 10.55776/PIN5424624. The author acknowledges TU Wien Bibliothek for financial support through its Open Access Funding Programme.
\end{acknowledgments}

\bibliography{svozil}
\ifws

\bibliographystyle{spmpsci}

\else

\fi

\end{document}


(* Function to compute the first n approximations (convergents) of the continued fraction expansion of Sqrt[2] *)
ContinuedFractionApproximationsSqrt2[n_Integer] := Module[
  {a, m = 0, d = 1, a0, h0 = 1, h1, k0 = 0, k1 = 1, h, k, approximations},

  a0 = Floor[Sqrt[2]];
  h1 = a0;
  approximations = {h1/k1}; (* The first approximation *)

  For[i = 1, i < n, i++,
    m = d*a0 - m;
    d = (2 - m^2)/d;
    a = Floor[(Sqrt[2] + m)/d];

    (* Update numerators and denominators for the next convergent *)
    h = a*h1 + h0;
    k = a*k1 + k0;

    (* Store the current convergent *)
    AppendTo[approximations, h/k];

    (* Update previous terms for the next iteration *)
    h0 = h1; h1 = h;
    k0 = k1; k1 = k;
    a0 = a;
  ];

  approximations
]

(* Example usage: compute the first 10 approximations *)
ContinuedFractionApproximationsSqrt2[5]


(* Function to compute the first n approximations of the binomial series expansion of Sqrt[2] *)
BinomialSeriesSqrt2[n_Integer] := Module[
  {approximation, term},

  approximation = 0;

  For[k = 0, k < n, k++,
    term = Binomial[1/2, k] * 1^k;
    approximation += term;
  ];

  approximation
]

(* Example usage: compute the first 10 approximations *)
Table[BinomialSeriesSqrt2[i], {i, 1, 5}]

RR:

Reviewer 1
This manuscript discusses the transition from unitary and reversible quantum dynamicsto non-unitary and irreversible measurement processes, using infinite tensor productsand nested Wigners Friend scenarios. The author shows how sectorization andfactorization can break unitary equivalence, and aims to derive the irreversibility ofobservation from this formalism. The argument is concise and theoretically ambitious,and the manuscript represents an interesting attempt to bring a fresh perspective to themeasurement problem.

Nevertheless, I believe the manuscript in its present form requires improvement before itcan be considered for publication. My main concerns are as follows:

(a)Sectorization and macroscopic states
The central claim is that sectorization explains irreversibility. However, the assumptionthat sectors can be identified with macroscopic systems or definite measurementoutcomes seems to presuppose precisely what should be explained. Mathematically,sectors are orthogonal subspaces of the infinite tensor product, but why they shouldcorrespond to pointer states or outcomes is not sufficiently justified. Without clarifyingthis, the claim that sectorization explains irreversibility seems to be circular.

(b) Contextuality of observations
The manuscript assumes that an observation or measurement amounts to ``adding acontext'' to the system, and that repeated observations trigger transitions betweensectors. Two important questions arises: (1) Why should observation be understood ascontext addition in the first place? (2) How is such a ``context'' physically realized, andwhy should it determine a sector? Without addressing these issues, the assumptionremains under-motivated. If this point is simply that observation or measurement leas tonon-unitary dynamics, then this adds little beyond the general claim that measurementsinduce non-unitary state-change.

(c) Clarification of the aim of this paper
It is not entirely clear what precise problem the paper intends to address. Is the aim toaccount for the emergence of irreversibility, the uniqueness of outcomes, or theinterpretative implications of the Wigner's friend scenario? The introduction would benefitfrom a clearer articulation of the target problem. In addition, the manuscript shouldexplain what the proposed formalism contributes in comparison to other approaches, andwhat the specific advantages of the framework are. If this issue is clarified, the need for asystematic engagement with previous literature, as discussed in (d), may beconsiderably reduced.

(d) Engagement with prior literature
The manuscript would benefit from a more systematic engagement with central strandsof prior work on the measurement problem, such as decoherence-based accounts,
collapse theories , and Everettian approaches. At present, the paper presents itsformalism largely in isolation, without critically situating it within these broader debates.Clarifying how the proposed account compares with, complements, or challenges theseestablished approaches would not only strengthen the philosophical relevance of thepaper, but also help the reader to understand its novelty and scope. If the issue raised in(c) is satisfactorily resolved, a detailed analysis of these debates may not be strictlynecessary.

(e) Relationship of Section IV with other sections
The other cases are discussed in Section IV -- drawn from real analysis and statisticalmechanics -- are intriguing, but their precise role in the argument remains unclear. Atpresent, they serve more as metaphor than as substantive evidence. While analogiescan help motivate intuitions, the paper should carefully distinguish between mathematicalirreversibility and physical irreversibility in measurement process. Otherwise there is arisk of overstating the relevance of these analogies. Are these analogies meant toprovide heuristic support, to demonstrate a structural parallel, or to establish explanatorylegitimacy? Clarifying this point would help the reader understand the function of thesesections.

(f) Minor point about representation.
On p.3, left column, about six lines from the bottom, the symbol ¥ket{¥down} isintroduced. For clarity , it would be helpful to add ¥ket{¥up} before ¥ket{down}.

In sum, the manuscript addresses an important topic and puts forward an intriguing approach. However, the current version relies too heavily on presuppositions (such assectors corresponding to macroscopic outcomes and observations as context addition)without sufficiently justifying them, does not clearly articulate the problem it seeks toresolve, and fails to engage systematically with central prior work on the measurementproblem. The analogies to other fields, while thought-provoking, do not yet serve theirintended purpose clearly, and may distract from the central claim unless their role is
clarified.

If the author can clarify the underlying assumptions, sharpen the articulation of thecentral problem, systematically situate the account with respect to existing approaches,and strengthen both the explanatory role of sectorization and the relevance of thehistorical analogies, this paper has the potential to make a valuable contribution to thephilosophy and foundations of quantum mechanics.

Reviewer 2

Report on
From Unitarity to Irreversibility: The Role of Infinite
Tensor Products and Nested Wigners Friends

The paper addresses the transition from reversible unitary quantum dynamics to irreversible
measurement processes, using the framework of infinite tensor products and nested Wigners
friend scenarios. The author argues that sectorization and factorization in infinite Hilbert
spaces can break unitary equivalence, offering a path to decoherence and irreversibility.
The paper provides a detailed analysis of how sectorization leads to orthogonal equivalence
classes of states, which cannot be coherently superposed across different sectors. Similarly,
factorization, analyzed within the theory of von Neumann algebras, introduces structural
changes that reinforce the loss of unitarity. Together, these mechanisms illustrate how macroscopic
irreversible behavior can be reconciled with the microscopic reversibility of quantum
mechanics. The discussion is enriched by historical and conceptual analogies, including connections
to Cantors diagonalization, computability limits (e.g., Specker sequences, Chaitins
halting probability), and Loschmidts paradox in statistical physics.
The manuscript is clearly written and well-referenced. While the reliance on infinite constructions
may limit direct operational applicability, the arguments are conceptually valuable.
However, a more explicit discussion of how the proposed mechanisms relate to standard decoherence
theory would further enhance the clarity and impact of the paper.
I recommend publication. The work provides a coherent and well-motivated contribution
to the foundations of quantum mechanics, particularly regarding the interplay between
unitarity, decoherence, and the emergence of classicality.

###################################################################################

\documentclass{article}
\usepackage[utf8]{inputenc}
\usepackage{amsmath}
\usepackage{geometry}
\geometry{a4paper, margin=1in}

\title{Response to Referee Reports}
\date{\today}

\begin{document}

\maketitle

\section*{General Response to the Referee Reports}

I kindly thank both referees for their diligent and insightful review of the manuscript. Their constructive comments were instrumental in strengthening the paper. Following their suggestions, I have revised the manuscript to clarify the central assumptions, explicitly situate the work within the context of existing literature, and sharpen the overall argumentation. I am confident that these changes have substantially improved the clarity and impact of the paper.

\section*{Point-by-Point Changes and Response to First Review}

Here is a detailed breakdown of the revisions made to the manuscript in response to each of the referee's comments.

\subsection*{(a) Sectorization and macroscopic states}

\noindent\textbf{Reviewer's Comment:}
\begin{quote}
The assumption that sectors correspond to macroscopic outcomes is not sufficiently justified and appears circular.
\end{quote}

\noindent\textbf{Change Made:}
A sentence was added to Section II.D to clarify that this connection is a postulate of the proposed framework.

\noindent\textbf{Before (from Section II.D):}
\begin{verbatim}
Therefore, for infinite tensor products, instead of directly dealing with the
entire infinite tensor product space,
one should consider regions or sectors within it. These sectors are equivalence
classes of vectors that differ only by a finite number of components
in the tensor product, or are otherwise close to (unitary equivalent) each other.
These sectors represent different `global' or `macroscopic'
configurations of the system~\cite{hepp-1972}.
\end{verbatim}

\noindent\textbf{After (from Section II.D):}
\begin{verbatim}
Therefore, for infinite tensor products, instead of directly dealing with the
entire infinite tensor product space,
one should consider regions or sectors within it. These sectors are equivalence
classes of vectors that differ only by a finite number of components
in the tensor product, or are otherwise close to (unitary equivalent) each other.
In this framework, it is postulated that these mathematically distinct sectors
correspond physically to different `global' or `macroscopic'
configurations of the system, such as pointer states of a measurement
apparatus~\cite{hepp-1972}.
\end{verbatim}

\subsection*{(b) Contextuality of observations}

\noindent\textbf{Reviewer's Comment:}
\begin{quote}
The concepts of ``context addition'' and its physical realization are under-motivated and unclear.
\end{quote}

\noindent\textbf{Change Made:}
The paragraph in Section III discussing this was rewritten to explicitly define ``context'' as the choice of measurement basis and explain how an infinite sequence of interactions (observations) drives the system into one of the newly formed sectors.

\noindent\textbf{Before (from Section III):}
\begin{verbatim}
According to their proposal, sectorization is a physical process in infinite
algebras: Separable
sectors correspond to `classical outcomes' and `macroscopic states' of
pointers~\cite{hepp-1972,bub-1988,bub-2015}.
Within a sector, the context and the resulting measurement have fixed macroscopic
values,
but a finite (sub)system within each sector can still undergo unitary evolution
as long as the state remains within its sector.
Any new measurement results in a shift to a different sector,
thus establishing a new context and corresponding outcome.
In the extreme case, a sector can be associated with a single element of a context.
Contexts and their associated sectors are `reshuffled' or `scrambled' when new
incompatible measurements are performed.
\end{verbatim}

\noindent\textbf{After (from Section III):}
\begin{verbatim}
According to their proposal, sectorization is a physical process in infinite
algebras where separable
sectors correspond to `classical outcomes' and `macroscopic states' of
pointers~\cite{hepp-1972,bub-1988,bub-2015}.
The ``context'' is physically realized by the choice of a measurement basis, which
is determined by the experimental arrangement. An ``observation'' then corresponds
to an interaction that entangles the system with the measurement apparatus,
projecting the state into this chosen basis. While a single such interaction is
a standard unitary evolution of the combined system, the formalism of infinite
tensor products suggests that an infinite sequence of such interactions forces
the total state into a new sector. This transition to a different, orthogonal
sector is what is meant by a ``shift'' that establishes a new macroscopic outcome.
A subsequent measurement in an incompatible basis (a new context) would similarly
drive the system into yet another distinct sector, a process described here as
`reshuffling' or `scrambling' of contexts.
\end{verbatim}

\subsection*{(c) \& (d) Clarification of the aim and Engagement with prior literature}

\noindent\textbf{Reviewer's Comment:}
\begin{quote}
The paper's aim is unclear, and it fails to engage with prior literature on the measurement problem (decoherence, collapse theories, Everettian approaches).
\end{quote}

\noindent\textbf{Change Made:}
A new paragraph was added to the Introduction to state the central problem, briefly mention the main existing approaches, and position this work as an alternative perspective. A corresponding paragraph was added to the Summary to explicitly compare this model with those other approaches.

\noindent\textbf{After (new paragraph in Section I, Introduction):}
\begin{verbatim}
The central problem this paper addresses is the quantum measurement problem:
how can the non-unitary, irreversible measurement process (von Neumann's
``process 1'') emerge from the purely unitary and reversible evolution described
by the Schr\"odinger equation (``process 2'')? Mainstream approaches to this problem
include decoherence-based accounts, which explain the apparent collapse as a
result of entanglement with the environment; objective collapse theories, which
modify quantum dynamics to include a physical collapse mechanism; and Everettian
or Many-Worlds interpretations, which posit that all outcomes occur in different
branches of a universal wavefunction. This paper explores an alternative
perspective, investigating the hypothesis that the transition to irreversibility
is an emergent phenomenon that arises in the mathematical limit of infinitely
complex systems. This limit is modeled using infinite tensor products, with the
nested Wigner's Friend scenario serving as a conceptual framework for such an
infinite regression. The advantage of this formalism is its ability to
mathematically break unitary equivalence without altering the fundamental quantum
postulates for finite systems.
\end{verbatim}

\noindent\textbf{After (new paragraph in Section V, Summary):}
\begin{verbatim}
The proposed mechanism can be contrasted with other solutions to the measurement
problem. Unlike objective collapse models, it does not postulate a new physical
collapse mechanism. In contrast to standard Everettian interpretations, the
sector structure implies that not all branches of the universal state are
equivalent, as transitions between them are forbidden by finite unitary means.
And while the model shares features with decoherence, which also relies on
interaction with a large environment, the use of infinite tensor products
provides a formal basis for a strict, rather than merely practical (FAPP),
breakdown of unitary equivalence between macroscopic outcomes.
\end{verbatim}

\subsection*{(e) Relationship of Section IV with other sections}

\noindent\textbf{Reviewer's Comment:}
\begin{quote}
The analogies in Section IV serve more as metaphor than evidence, and their role in the argument is unclear.
\end{quote}

\noindent\textbf{Change Made:}
An introductory sentence was added to Section IV to explicitly state that the analogies are intended as heuristic support to motivate the central idea, not as direct proof.

\noindent\textbf{Before (Section IV):}
\begin{verbatim}
\section{Historical Analogues}

This section explores several related but distinct concepts that have been
investigated in various areas of physics.
\end{verbatim}

\noindent\textbf{After (Section IV):}
\begin{verbatim}
\section{Historical Analogues}

To motivate the use of infinite limits as a tool for explaining emergent
irreversibility, this section presents analogies from other fields of
mathematics and physics where infinite processes lead to qualitatively new
phenomena not present in their finite counterparts. These examples are intended
as heuristic support to demonstrate a recurring structural parallel, rather
than as direct physical evidence for the quantum mechanical argument.
This section explores several related but distinct concepts that have been
investigated in various areas of physics.
\end{verbatim}

\subsection*{(f) Minor point about representation}

\noindent\textbf{Reviewer's Comment:}
\begin{quote}
For clarity, it would be helpful to introduce `\textbackslash ket\{\textbackslash up\}' before `\textbackslash ket\{\textbackslash down\}'.
\end{quote}

\noindent\textbf{Change Made:}
The text in Section II.C.3 was slightly modified to explicitly reference the previously defined `\(\vert \uparrow \rangle\)' state when introducing the orthogonal `\(\vert \downarrow \rangle\)' state.

\noindent\textbf{Before (from Section II.C.3):}
\begin{verbatim}
On the other hand, for any vector
containing a component orthogonal to \(\begin{pmatrix} 1 , 0 \end{pmatrix}^\intercal\)
in at least one factor
$
\vert \downarrow \rangle = \begin{pmatrix}  0,1  \end{pmatrix}^\intercal
$,
\(F\) maps it to the zero vector.
\end{verbatim}

\noindent\textbf{After (from Section II.C.3):}
\begin{verbatim}
On the other hand, for any vector
containing a component orthogonal to \(\vert \uparrow \rangle = \begin{pmatrix} 1 , 0 \end{pmatrix}^\intercal\), such as the
spin-down state $ \vert \downarrow \rangle =
\begin{pmatrix}  0,1  \end{pmatrix}^\intercal $, in at least one factor,
\(F\) maps it to the zero vector.
\end{verbatim}

\section*{Point-by-Point Changes and Response to Second Review}

\noindent\textbf{Reviewer's Comment:}
\begin{quote}
"A more explicit discussion of how the proposed mechanisms relate to standard decoherence theory would further enhance the clarity and impact of the paper.''
\end{quote}

\noindent\textbf{Change Made:}
In response to this excellent suggestion, a new paragraph has been added to the \texttt{Summary} section. This paragraph explicitly compares and contrasts the formalism presented in the paper with standard decoherence theory, focusing on the distinction between practical (FAPP) and formal irreversibility.

\noindent\textbf{Text added to Section V (Summary):}
\begin{verbatim}
The framework presented here also shares significant conceptual ground with
standard decoherence theory, yet it is distinguished by a crucial formal
difference. Both approaches attribute the loss of quantum coherence to the
system's entanglement with a larger, more complex entity---be it a physical
environment or, in this paper's model, an infinite chain of observers.
The selection of a preferred basis (the 'pointer basis' in decoherence) is
analogous to the formation of sectors corresponding to macroscopic outcomes.
The primary divergence, however, lies in the nature of the resulting
irreversibility. Standard decoherence describes a practical (FAPP) process
within a finite, albeit large, system-environment composite. The global
evolution remains unitary, and coherence is merely 'leaked' into the
environmental degrees of freedom, becoming locally inaccessible but never
truly destroyed. In contrast, the mechanism proposed here, leveraging the
mathematical properties of the infinite tensor product limit, describes a
formal and absolute breakdown of unitary equivalence. The emergence of sectors
is not a matter of information being difficult to retrieve; it is a structural
feature of the infinite-dimensional Hilbert space where states in different
sectors are mathematically non-interconvertible by finite unitary means. Thus,
while decoherence explains the \emph{appearance} of classicality in a
fundamentally unitary world, this paper's formalism offers a mathematical
route to the \emph{emergence} of genuine, non-unitary irreversibility in the
thermodynamic limit.
\end{verbatim}

\end{document}